\documentclass[aps,prl,reprint,groupedaddress]{revtex4-1}

\usepackage{graphicx}
\usepackage{dcolumn}
\usepackage{inputenc}

\begin{document}

\title{Effect of a Physical Phase Plate on Contrast Transfer in an Aberration-Corrected Transmission Electron Microscope}

\author{B. Gamm}
\email{gamm@kit.edu}
\affiliation{Laboratorium f\"ur Elektronenmikroskopie, Karlsruher Institut f\"ur Technologie (KIT), 76131 Karlsruhe, Germany}
\author{K. Schultheiss}
\affiliation{Laboratorium f\"ur Elektronenmikroskopie, Karlsruher Institut f\"ur Technologie (KIT), 76131 Karlsruhe, Germany}

\author{D. Gerthsen}
\affiliation{Laboratorium f\"ur Elektronenmikroskopie, Karlsruher Institut f\"ur Technologie (KIT), 76131 Karlsruhe, Germany}
\author{R.R. Schr\"oder}
\affiliation{Max-Planck-Institute of Biophysics, Max-von-Laue-Str. 3, D-60438 Frankfurt am Main, Germany}

\date{\today}

\begin{abstract}
In this theoretical study we analyze contrast transfer of weak-phase objects in a transmission electron microscope, which is equipped with an aberration corrector (C$_{S}$-corrector) in the imaging lens system and a physical phase plate in the back focal plane of the objective lens. For a phase shift of $\Pi/2$ between scattered and unscattered electrons induced by a physical phase plate, the sine-type phase contrast transfer function is converted into a cosine-type function. Optimal imaging conditions could theoretically be achieved if the phase shifts caused by the objective lens defocus and lens aberrations would be equal zero. In reality this situation is difficult to realize because of residual aberrations and varying, non-zero local defocus values, which in general result from an uneven sample surface topography. We explore the conditions - i.e. range of C$_{S}$-values and defocus - for most favourable contrast transfer as a function of the information limit, which is only limited by the effect of partial coherence of the electron wave in C$_{S}$-corrected transmission electron microscopes. Under high-resolution operation conditions we find that a physical phase plate improves strongly low- and medium-resolution object contrast, while improving tolerance to defocus and C$_{S}$-variations, compared to a microscope without a phase plate.
\end{abstract}

\pacs{87.64.Ee,87.64.mh,41.85.Ne}

\maketitle

\section{Introduction}
The resolution and interpretability of transmission electron microscopy (TEM) images have been hampered for decades by the strong aberrations of magnetic electron lenses. In particular the spherical aberration has been the dominant resolution-limiting aberration. The most severe consequence of spherical aberration is the strong dependence of the induced phase shifts on the spatial frequency, which results in the delocalization of image information and therefore impedes intuitive interpretability of high-resolution TEM images. Based on the idea of Scherzer \cite{Scherzer1947} and the theoretical concept of Rose \cite{Rose1990}, Haider et al. \cite{Haider1998} realized the first double-hexapole aberration corrector – for short C$_{S}$-corrector - which allows correction of the spherical aberration as well as other aberrations. With the successful implementation of the C$_{S}$-corrector a new generation of transmission electron microscopes with unprecedented resolution and imaging capabilities has meanwhile emerged. 
With the availability of the C$_{S}$-corrector, the spherical aberration coefficient C$_{S}$ can be considered as a second tuneable parameter in addition to the objective lens defocus to optimize image contrast and resolution of the microscope. Lentzen et al. \cite{Lentzen2002} recognized that it is not useful to shift the first zero of the phase contrast transfer function significantly beyond the information limit and derived an optimum combination of C$_{S}$-value and defocus at which the point resolution for phase-contrast imaging is close to the information limit. These parameters can be considered to yield optimum contrast in a sense that a large interval of spatial frequencies is transmitted with a large phase shift without oscillation of the contrast transfer function. 
The imaging conditions of aberration-corrected imaging are very often optimized to obtain highest resolution, generally at the cost of strong phase contrast information at intermediate and low spatial frequencies. Important alternative techniques for imaging at highest resolution without diminishing object contrast at lower resolution are off-axis electron holography \cite{Tonomura1982,Lichte1986} and through focus series \cite{Coene1996,Thust1996}. These techniques require special experimental arrangements and additional numerical processing of image data to retrieve the true projected phase shifting potential of the object, however, these techniques have been applied most successfully over the recent years. Holography has the additional advantage to allow quantitative imaging, i.e. the direct measurement of the relative phase shifts is possible.

A rather novel possibility to image objects at highest resolution and optimal phase contrast over the whole resolution range has been made available most recently by the development of physical phase plates. Such phase shifting devices can be placed in the back focal plane of the objective lens or any other conjugate diffraction plane. In analogy to Zernike´s $\pi/4$ phase plate in light microscopy \cite{Zernike1934}, a relative phase shift of $90^\circ$ between scattered and axial electrons leads then to a substantial contrast enhancement in TEM images. Several concepts for physical phase plates for transmission electron microscopes were realized recently. The Zernike-type phase plate of Danev and Nagayama \cite{Danev2001} consists of a thin amorphous carbon film with a small hole in the center where the unscattered electrons propagate without additional phase shift, whereas the phase shift of the scattered electrons is determined by the thickness of the carbon film. An electrostatic Boersch phase plate \cite{Boersch1947} was fabricated by our group \cite{Schultheiss2006} and also by Huang et al. \cite{Huang2006}, in both cases implementing designs of Matsumoto and Tonomura \cite{Matsumoto1996} and Majorovits and Schröder \cite{EMajorovits}. The electrostatic drift-tube phase plate of Cambie et al. \cite{Cambie2007} is based on a different, two-electrode design. For weak-phase objects, optimum phase contrast is achieved for a relative phase shift $\phi_{pp} = \pi/2$ between scattered and axial electrons. For the electrostatic Boersch phase plate and the electrostatic drift tube the applied electric potential can be considered as another tuneable parameter for optimizing image information. 
It was shown in a theoretical study by Lentzen \cite{Lentzen2004} that a  C$_{S}$-corrector could be tuned to approximate a Zernike-type phase plate where a constant phase shift for a maximum interval of spatial frequencies can be obtained to optimize high-resolution phase contrast. However, only small phase shifts are achieved for small  C$_{S}$-values at low and intermediate spatial frequencies, which results in vanishing phase contrast for large- and medium-scale object features of weak-phase objects. The frequency dependence of contrast of a "Lentzen phase plate" is thus completely different from the uniform contrast transfer of a physical phase plate, which therefore may prove to be the ideal electron-optical device to improve contrast at any given resolution.
In the present work we discuss the prospects to optimize contrast transfer by combining Cs-correction and a physical phase plate in a transmission electron microscope. Two “model” microscopes characterized by information limits of 0.12 nm and 0.05 nm will be considered. We show that the imaging conditions for phase contrast obtained by applying a physical phase plate are relating defocus and residual lens aberration in a similar way as for more conventional imaging. It will become obvious that contrast transfer is significantly improved by the implementation of a phase plate. We show that this is the case for both, the idealised field-only phase plate and our current miniaturized structure of the physical phase plate. We will also analyse the robustness of high-resolution imaging conditions with respect to variations of the objective lens defocus and  C$_{S}$. 

\section{Phase Contrast Theory}

The following discussion is based on the concept of the phase contrast transfer function (PCTF), which is commonly used to describe contrast transfer for weak-phase objects and resolution in transmission electron microscopy. The coherent PCTF is given by sin($\chi$) which contains the aberration function 
\begin{equation}\label{eqn1}
\chi = \pi ( Z \lambda u^2 + \frac{1}{2} C_{S} \lambda^3 u^4
\end{equation}
with Z denoting the objective lens defocus, u the spatial frequency and $\lambda$ the electron wavelength. Scherzer \cite{Scherzer1949} derived an optimum defocus  , which allows transmission of a wide interval of spatial frequencies with a nearly constant phase shift. The resolution for $Z_{Sch}$, i.e. the 1st zero of the phase contrast transfer function, is characterized by a spatial frequency $u_{Sch} = 1.52 (C_{S} \lambda^3)^{-1/4}$.
Contrast transfer in TEM is limited by the partial coherence of the electron wave, which can be described by envelope functions for partial temporal coherence $E_{t}$ and partial spatial coherence $E_{S}$ given by Eqs. (\ref{eqn2},\ref{eqn3})
\begin{equation}\label{eqn2}
E_{t} = exp(-\frac{1}{2}(\pi \lambda \Delta_{Z})^2 u^4)
\end{equation}
\begin{equation}\label{eqn3}
E_{S}=exp[-(\frac{\pi \alpha}{\lambda}) (C_{S} \lambda^3 u^3 + Z \lambda u)^2]
\end{equation}

The information limit is defined by the spatial frequency, where the product $E_{S} \cdot  E_{t}$ falls below the noise level given by $e^{-1/2} = 0.136$ . The calculations in the following are performed for a 200 keV transmission electron microscope with an information limit of 0.12 nm (denoted as microscope 1) and a 300 keV microscope (microscope 2) with an information limit of 0.05 nm. Microscope 1 can be considered as a state-of-the-art C$_{S}$-corrected microscope whereas microscope 2 corresponds to a high-end microscope, which may be nevertheless available in the near future. To adjust the information limit, the defocus spread $\Delta_{Z}$ was set at 4 nm for microscope 1 and 0.8 nm for microscope 2. An illumination angle $\alpha$ of 1 mrad was chosen.

\subsection{Transmission Electron Microscope with C$_{S}$-Corrector}

For a C$_{S}$-corrected transmission electron microscope Scherzer resolution is only limited by the information limit of the instrument. However, Lentzen et al. \cite{Lentzen2002} recognized that it is not useful to extend the first zero of the PCTF significantly beyond the information limit because extremely small C$_{S}$-values lead to a large interval with vanishing phase contrast at small and intermediate spatial frequencies. They derived optimum C$_{S}$- and Z-values for high phase contrast and low delocalization by equating Scherzer and Lichte defocus  $Z_{L} = -3/4 C_{S} \lambda^2 u_{inf}^2$. The optimum values C$_{S,opt}$ and Z$_{opt}$ are given by Eqs.(\ref{eqn4},\ref{eqn5}):
\begin{equation}\label{eqn4}
C_{S,opt}=\frac{64}{27} \frac{1}{\lambda^3 u_{inf}^4}
\end{equation}
\begin{equation}\label{eqn5}
Z_{opt}=-\frac{16}{9} \frac{1}{\lambda u_{inf}^2}
\end{equation}

Moreover, Jia et al. \cite{Jia2004} showed that negative C$_{S,opt}$ values with the corresponding positive Z$_{opt}$ lead to a contrast reversal of atoms from dark to bright in thin specimens and a strong contrast enhancement compared to positive C$_{S}$ values. The bright-atom contrast results from the sign reversal of the phase shift imposed by the objective. Image simulations demonstrate that the strong contrast enhancement for negative C$_{S}$-values is correlated with the effects of nonlinear image formation.

\subsection{Transmission Electron Microscope with a Physical Phase Plate}
In the following, contrast transfer for a microscope with a physical phase plate inserted in the back focal plane - but without C$_{S}$-corrector - is considered. The aberration function $\chi$ is now complemented by an additional phase shift $\phi_{pp}$ induced by the phase plate.
\begin{equation}\label{eqn6}
\chi = \pi ( Z \lambda u^2 + \frac{1}{2} C_{S} \lambda^3 u^4 + \phi_{pp}
\end{equation}
Assuming an additional phase shift $\phi_{pp} = \pi/2 $ converts the sine-type PCTF into a cosine-type function. This results in an improved contrast transfer at low and medium spatial frequencies. Danev and Nagayama \cite{Danev2001} calculated the optimum defocus for $\phi_{pp} = \pi/2$ given by Eq.(\ref{eqn7})
\begin{equation}\label{eqn7}
Z_{Sch,pp}=-0.73 \sqrt{C_{S} \lambda}
\end{equation}
using the same criteria as Scherzer. The corresponding Scherzer resolution is given by Eq.(\ref{eqn8})
\begin{equation}\label{eqn8}
u_{Sch,pp}=1.4 (C_{S} \lambda^3)^{-1/4}
\end{equation} 
which is slightly worse than the original Scherzer resolution $u_{Sch}$. 

\subsection{Transmission Electron Microscope with a Physical Phase Plate and C$_{S}$-Corrector}
The considerations in this section apply to a transmission electron microscope with C$_{S}$-corrector, which is also equipped with a phase plate in the back focal plane of the objective lens. We assume an ideal phase plate without obstructing bars, which are nevertheless part of the design of electrostatic phase plates \cite{Schultheiss2006,Huang2006,EMajorovits,Cambie2007}, or coherence loss due to inelastic scattering in the case of a Zernike-type phase plate \cite{Danev2001}. 

\begin{figure}
\includegraphics[width=0.45\textwidth]{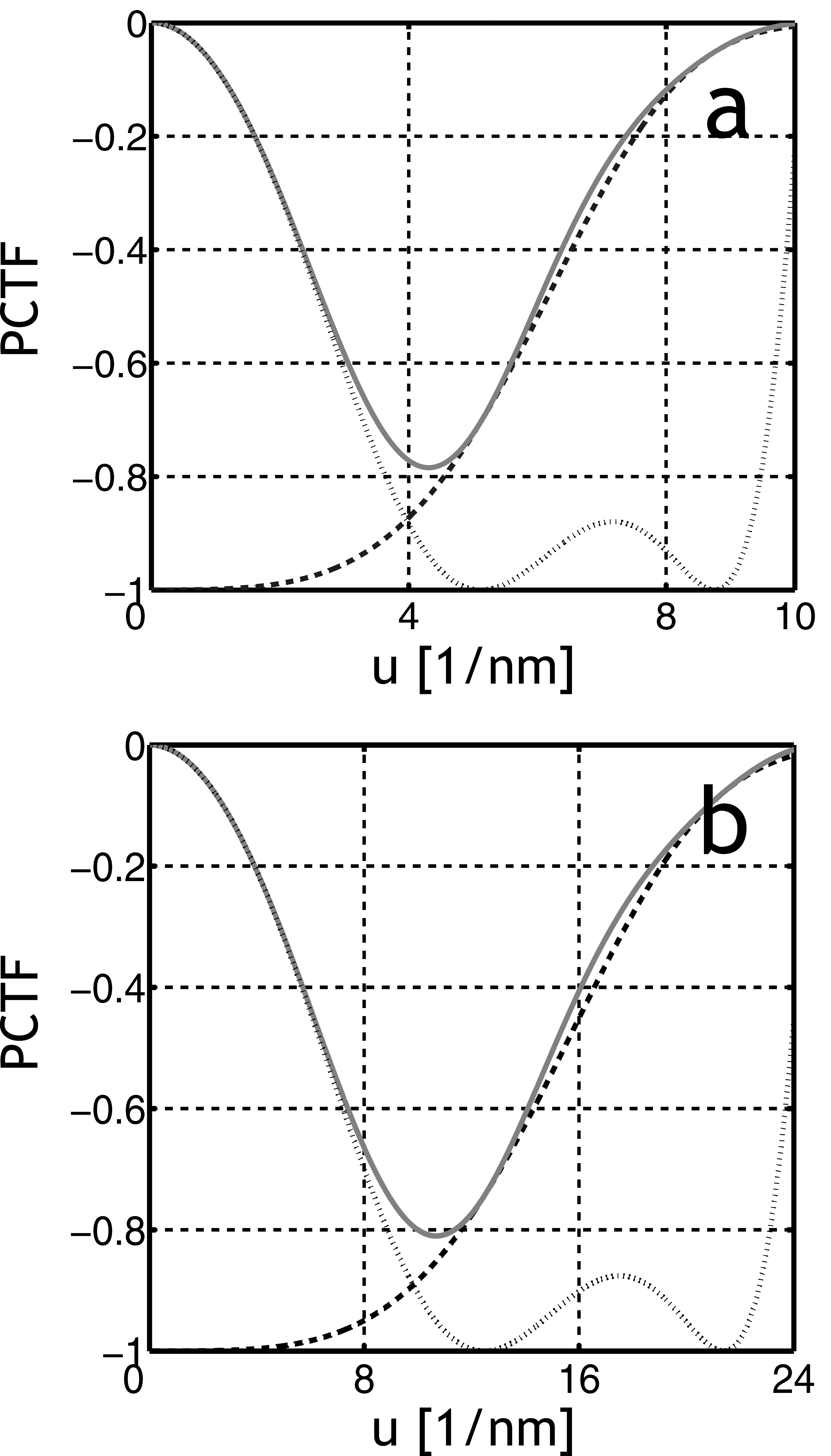}
\caption{\label{fig:fig1}Phase-Contrast-Transfer-Function (PCTF, solid gray line), sine-function of wave aberration $sin(\chi)$ (dotted line) and envelope function (dashed black line) for a) 200 keV microscope with an information limit of 0.12 nm and b) 300keV microscope with an information limit of 0.05 nm.}
\end{figure}
\begin{figure*}
\includegraphics[width=0.45\textwidth]{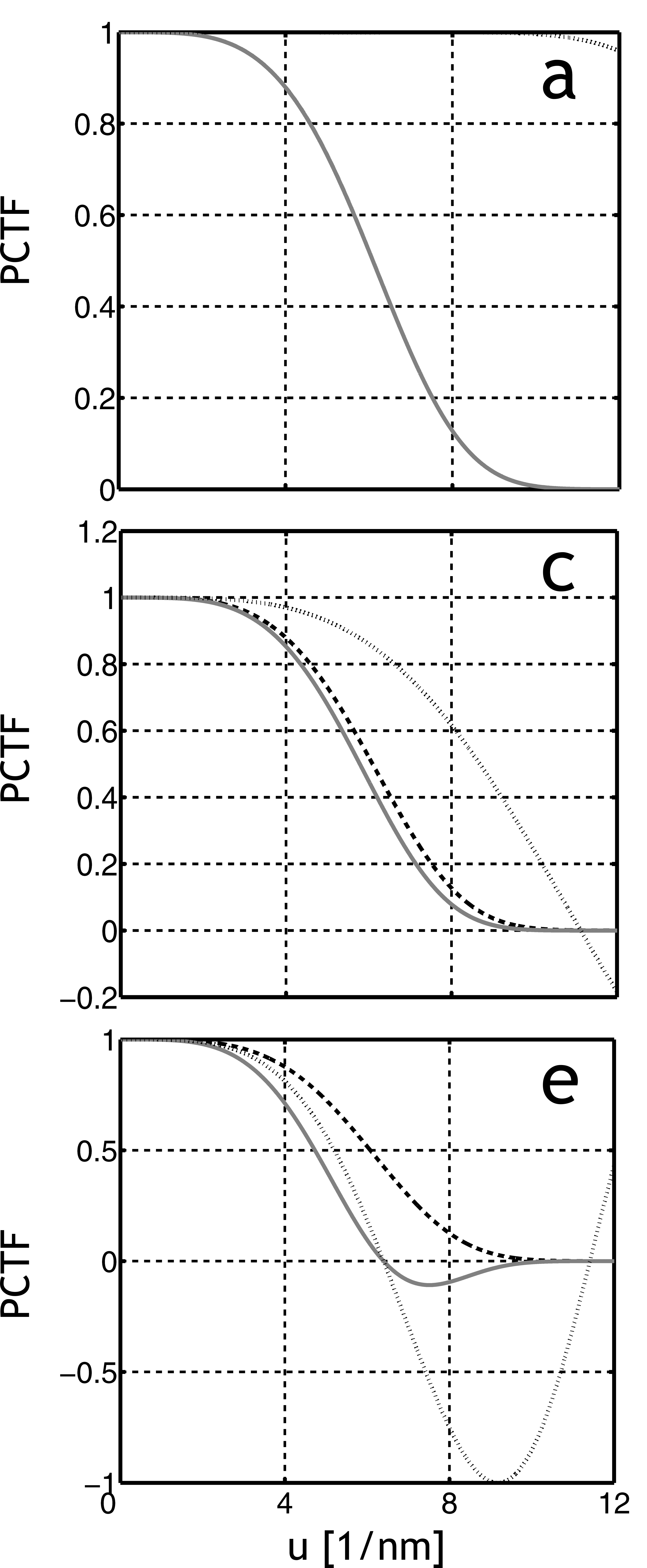}\includegraphics[width=0.45\textwidth]{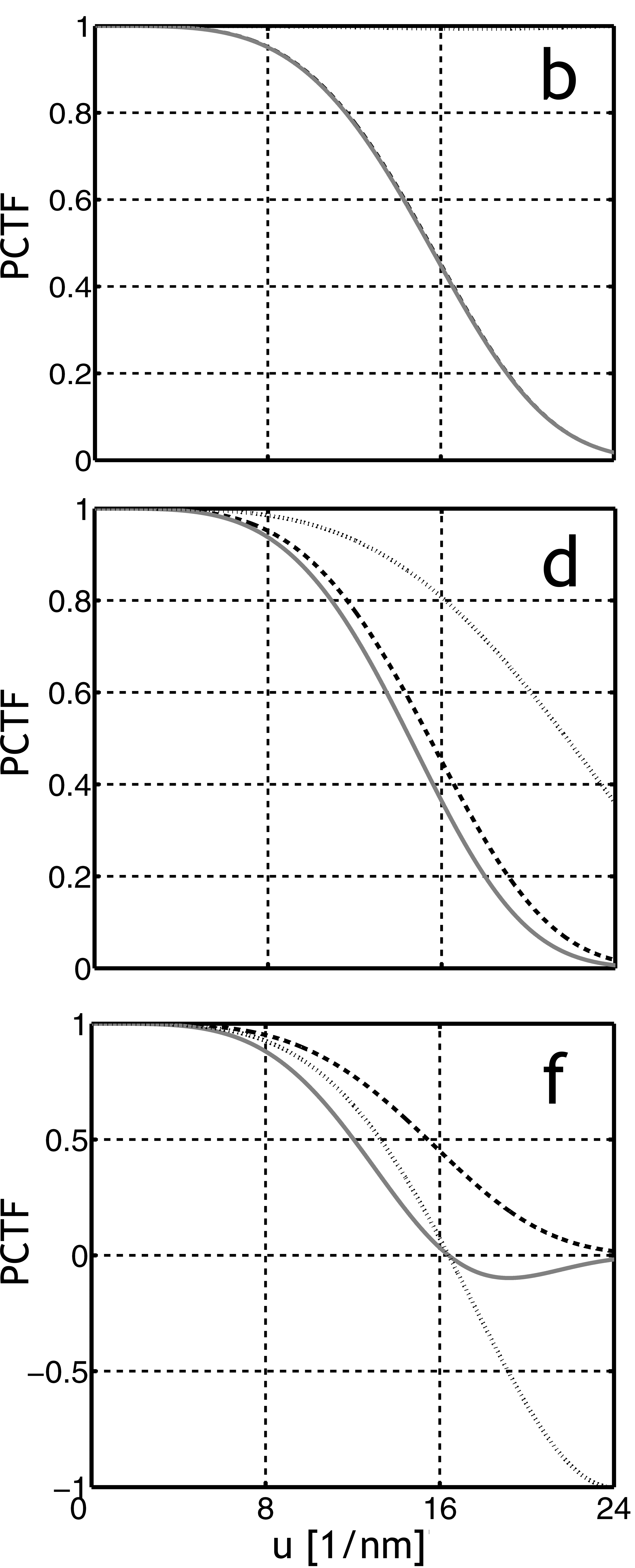}
\caption{\label{fig:fig2}Phase-Contrast-Transfer-Function (PCTF, solid gray line), sine-function of wave aberration $sin(\chi$) (dotted line) and envelope function for a 200 keV microscope with an information limit of 0.12 nm (microscope 1) for $C_{S} = 1~\mu m$ and a defocus value of a) Z = -0.2 nm, c) Z = -2.0 nm and e) Z = -5 nm and for a 300 keV microscope with an information limit of 0.05 nm (microscope 2) for $C_{S} = 0.1~\mu m$ and b) Z = -0.12 nm, d) Z = – 0.45 nm and f) Z = - 1 nm.}
\end{figure*}

To illustrate the improvement of contrast transfer, PCTFs without and with an ideal phase plate are plotted in Fig. \ref{fig:fig1} and Fig. \ref{fig:fig2}. The PCTF for microscope 1 without phase plate in Fig. \ref{fig:fig1}(a) is calculated with C$_{S,opt}$ = 36.5 $\mu m$ and Z$_{opt}$ = - 11.1 nm according to Eqs.(\ref{eqn4},\ref{eqn5}). The corresponding parameters for microscope 2 (Fig. \ref{fig:fig1}(b)) are C$_ {S,opt}$ = 1.9 $\mu m$ and Z$_{opt}$ = - 2.2 nm. Figs. \ref{fig:fig1}(a,b) contain the coherent PCTF (dotted lines) and the product $E_{t} \cdot E_{S}$ (bold black lines) in addition to the PCTF (dashed gray lines). The PCTFs for the C$_{S}$-corrected instruments coincide with the damping envelopes from intermediate to high spatial frequencies but it decreases towards zero at lower spatial frequencies. In contrast, perfect coincidence of the PCTF and the envelope functions can be realized for a microscope with a phase plate - independent of the particular instrument - for the whole spectrum up to the information limit by setting C$_{S}$ = 0 $\mu m$, Z = 0 nm and $\phi_{pp}= \pi/2$ which implies $sin(\phi_{pp}) = 1$.

However, these ideal conditions can be adjusted only with a finite accuracy, which depends on the measurement precision for C$_{S}$, Z and $\phi_{pp}$. A further limitation is given by the topography of the exit surface of the sample, which determines the local defocus value. As a consequence, deviations of the spherical aberration coefficient $\Delta C_{S}$ and defocuses $\Delta Z$ from the ideal values have to be analyzed with respect to maintaining optimum contrast transfer. We note already here that perfect coincidence of the PCTF and the envelope functions can be only obtained for very small C$_{S}$ values. 
The following considerations yield estimates for Z, up to which (almost) perfect coincidence of the PCTF with the envelope functions can be obtained if the C$_{S}$ values are small enough. For a phase shift $\phi_{pp} = \pi/2$ the first two maxima of the PCTF are situated at $u_{1} = 0 nm^{-1}$ and $u_{2}=\sqrt{-\frac{2 Z}{C_{S} \lambda^2}}$ with a local minimum at $u_{min} = \sqrt{-\frac{Z}{C_{S} \lambda^2}}$ in between, if Z and Cs have opposite sign. Optimized contrast transfer can be expected for the defocus value Z$_{1}$ given by Eq.(\ref{eqn9}) where the 2nd maximum of the PCTF coincides with the information limit
\begin{equation}\label{eqn9}
Z_{1}=- \frac{1}{2}  C_{S} \lambda^2 u_{inf}^{2}
\end{equation}

\begin{figure*}
\includegraphics[width=0.7\textwidth]{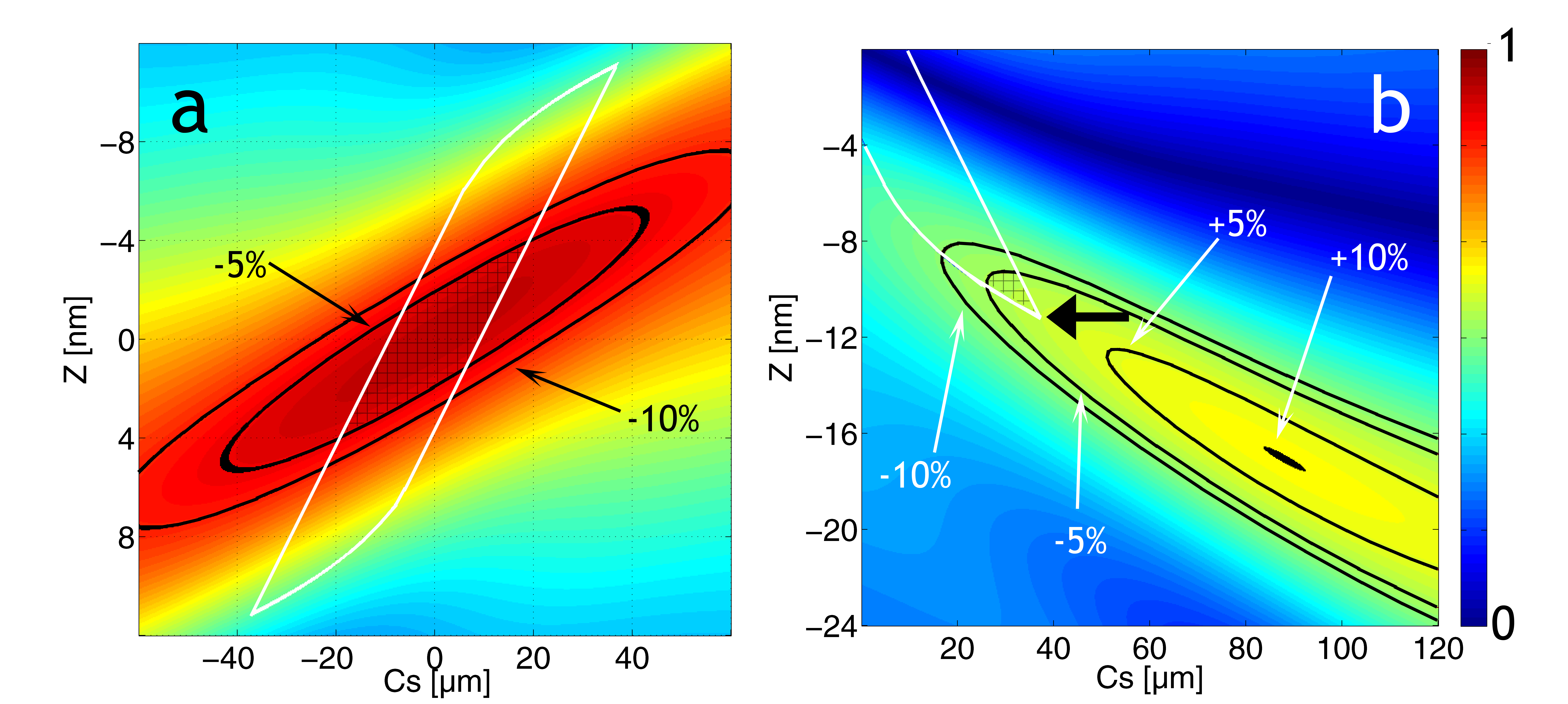}\hfill
\includegraphics[width=0.7\textwidth]{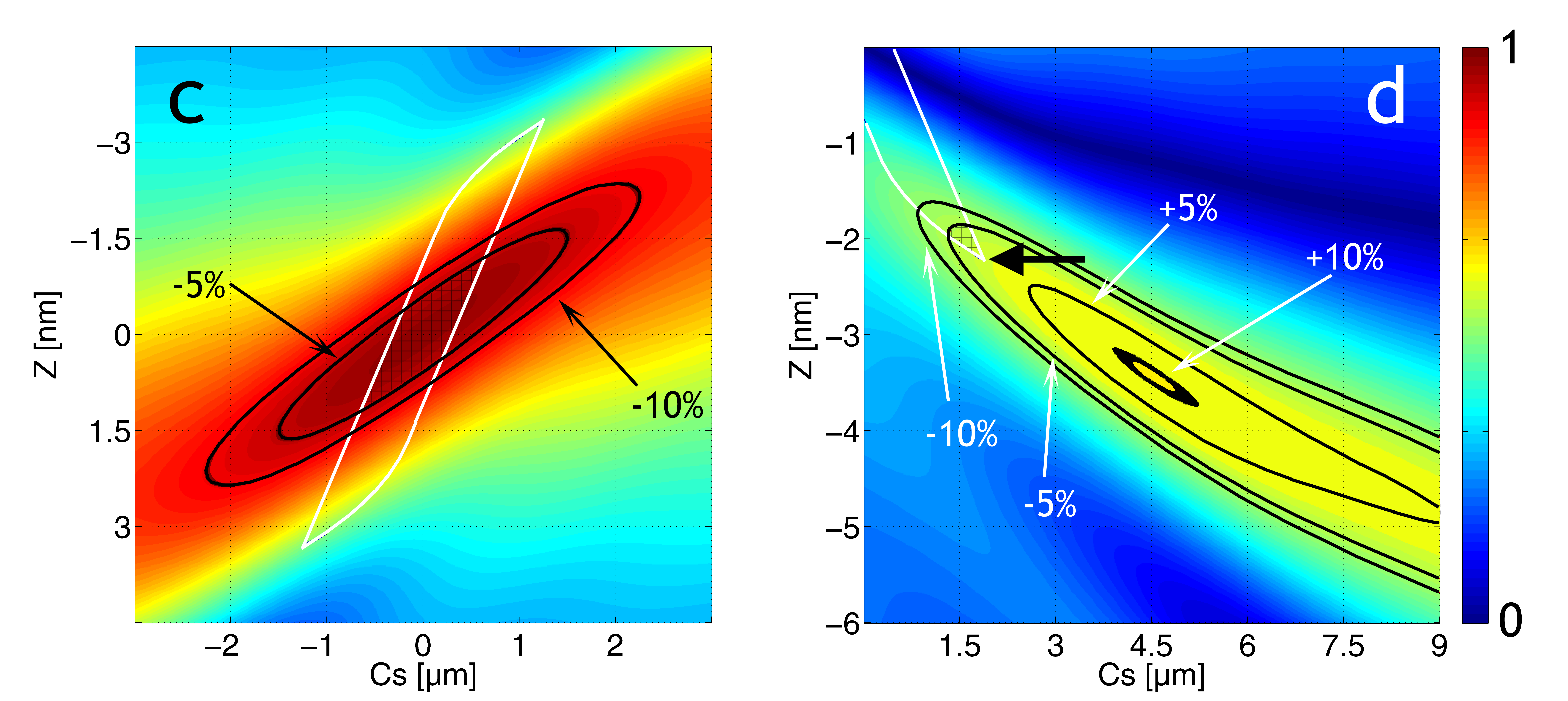}
\caption{\label{fig:fig3}Color-coded plot of the integrated phase contrast transfer PC (Eq. (\ref{eqn10})) normalized with respect to the ideal value PC$_{id}$ (Eq. (\ref{eqn11})) for a) microscope 1 (information limit 0.12 nm) with a phase plate and b) without phase plate and for microscope 2 (information limit 0.05 nm) c) with phase plate and d) without phase plate. The bold black lines denote the parameter ranges with 0.95 and 0.9 PC$_{id}$ in a) and c). In b) and d) the black lines indicate parameter ranges with 0.9, 0.95, 1.05 and 1.1 PC$_{len}$. The white lines in all figures encircles parameters with optimized delocalization of the object information R<R$_{len}$.}
\end{figure*}

In the following we assume an adjustment precision $\Delta C_{S} = \pm 1 \mu m$ for microscope 1 which can be well achieved by today's C$_{S}$-correctors \cite{Haider}. For the high-end microscope 2 we adopt $\Delta C_{S} = \pm 0.1 \mu m$. Setting accordingly $C_{S} = 1 \mu m$ for microscope 1 and $C_{S} = 0.1 \mu m$ for microscope 2 to the limit of these adjustment tolerances, the PCTFs are plotted in Fig. \ref{fig:fig2}(a,b) for microscope 1 with $Z_{1}= - 0.2 nm$ and microscope 2 with $Z_{1}= - 0.12 nm$. Under these conditions, the deviation between the PCTF and envelope functions is indeed negligible. Although the defocus values $Z_{1}$ may be too small to be set precisely (also due to variations of the exit surface topography of the sample), a certain interval of defocus values $\Delta Z$ can be admitted for which almost perfect conditions are maintained. The width of this interval depends on $C_{S}$ and the information limit of the microscope as will be demonstrated later in Fig. \ref{fig:fig3} and Fig. \ref{fig:fig4}. The effect of larger Z on the PCTF is illustrated in Figs. \ref{fig:fig2}(c,e) for microscope 1 with Z = -2.0 nm and Z = - 5 nm and in Figs. \ref{fig:fig2}(d,f) with Z = -0.45 nm and Z= -1.0 nm for microscope 2. The deviation of the PCTFs from the envelope functions is still relatively small but we observe a zero transition of the PCTF for the largest underfocus values in Figs. \ref{fig:fig2}(e,f) and correspondingly negative values of PCTF at higher spatial frequency u. This corresponds to a reversal of the sign of the phase shift imposed by the objective lens, which is undesirable because information from a weak-phase object within this interval of spatial frequencies would be visible with reversed (dark) contrast in the image as opposed to bright contrast for information with smaller spatial frequencies.

As a measure for phase contrast transfer we introduce the integral of the PCTF up to the information limit 
\begin{equation}\label{eqn10}
PC=\int_{0}^{u_{inf}} \! E_{t} \cdot E_{S} \left| sin(\chi_{pp}) \right| du
\end{equation}
We note that the integration is only carried out up to the first zero of the PCTF if it is situated at smaller spatial frequencies than the information limit to avoid contrast reversal. The value of the integral given by Eq. (\ref{eqn10}) is compared to the ideal case for a microscope with phase plate (PP), which is represented by
\begin{equation}\label{eqn11}
PC_{id}=\int_{0}^{u_{inf}} \! E_{t} \cdot E_{S} du
\end{equation}
To estimate the range of defocus values $\Delta Z$ and $\Delta C_{S}$ for high phase contrast, we consider the reduction of PC with respect to PCid, which is defined by $\frac{PC_{id} - PC}{PC_{id}}$. Examples of PCTFs for a 5 \% reduction of PC with respect to $PC_{id}$ are depicted in Figs. 2(c,d). For a microscope without PP, where PC never reaches $PC_{id}$, the deviation of the phase contrast is measured with respect to the value of PC at the Lentzen parameters given by Eqs. (\ref{eqn4},\ref{eqn5}) denoted as $PC_{len}$, i.e. $\frac{PC_{len} - PC}{PC_len}$.
Apart from phase contrast transfer, delocalisation is another important property to consider for the image quality. The delocalisation R is given by the maximum of the derivative of $\chi$ \cite{Lichte1991}  
\begin{equation}\label{eqn12}
R=max(\frac{\partial \chi}{\partial u})
\end{equation}
It is apparent that R has the same dependence on C$_{S}$ and Z for a PP microscope and a non-PP microscope, because the derivatives of $\chi$ are in both cases the same.

The colour-coded plots in Fig. \ref{fig:fig3} show the PC values for microscope 1 with (Fig. \ref{fig:fig3}(a)) and without PP (Fig. \ref{fig:fig3}(b)) as well as for microscope 2 with (Fig. \ref{fig:fig3}(c)) and without PP (Fig. \ref{fig:fig3}(d)). The colour scale is identical in plots for the same microscope to visualize the differences of PC with and without PP. As expected, the PC values for microscopes without PP reach only about 60 \% of $PC_{id}$. We note already here that the loss of information due to the electrode structure amounts to at most 8 \% for microscope 1 and 3 \% for microscope 2 depending on the objective lens properties (focal length) and the electron wavelength, which still leaves considerably stronger phase contrast for a PP microscope. The parameter ranges encircled by the black lines in Figs. \ref{fig:fig3}(a,c) mark C$_{S}$- and Z-values with a 5 \% and 10 \% reduction with respect to $PC_{id}$. For the non-PP microscopes, the optimal Lentzen-parameter settings and its corresponding $PC_{len}$, indicated by the arrows in Figs. \ref{fig:fig3}(b,d), do not represent the maximum achievable PC value. Therefore, the black lines in Figs. \ref{fig:fig3}(b,d) mark parameters with 0.9, 0.95, 1.05 and 1.1 $PC_{len}$. The white lines denote the parameters for a delocalisation, which is expected for Lentzen parameters in a non-PP microscope. Fig. \ref{fig:fig3}(a) shows that a large range of C$_{S}$-values and $\Delta Z \approx 4 nm$ for only 5 \% reduction with respect to $PC_{id}$ and delocalisation smaller than R$_{len}$. $\Delta C_{S}$ and $\Delta Z$ are symmetrical with respect to the origin of the coordinate system for $\phi_{pp} = \pi/2$. Fig. \ref{fig:fig3}(b) shows the case for a non-PP microscope in the sector for negative Z and positive C$_{S}$. The plot is symmetrical to positive Z and negative C$_{S}$ values. If we allow a 5 \% reduction of PC with respect to $PC_{len}$ and require the delocalisation to be at least as small as $R_{len}$, the area (shaded areas in Fig. \ref{fig:fig3}) enclosed by the intersecting  white and black lines define the range of suitable microscope parameters. The maximum defocus range here is only $\Delta Z \approx 1.4 nm$. As expected, the tolerances for microscope 2 are far more stringent both with (Fig. \ref{fig:fig3}(c)) and without PP (Fig. \ref{fig:fig3}(d)). We emphasize here another benefit of a PP microscope: The minimum delocalization of image information coincides with optimized conditions around $C_{S} = 0~ \mu m$ and Z = 0 nm.

Since only few objects are pure weak-phase objects we finally discuss objects, which also provide amplitude contrast. Phase contrast in a PP microscope behaves like amplitude contrast in a non-PP instrument and vice versa. For $\phi_{pp} = + \pi/2$ amplitude contrast is transferred with a sin-type transfer function which means suppression of amplitude information with low and intermediate spatial frequencies. To obtain both amplitude and phase contrast, a pair of images with the same defocus Z and $\phi_{pp} = + \pi/2$ and $\phi_{pp} = 0$ can be recorded from which the image can be reconstructed by complex reconstruction \cite{Nagayama1999}. The reconstruction algorithm can be described schematically in the following way: The Fourier-transformed images with phase plate F(I$_{pp}$) and without PP (F(I$_{bright}$)) correspond to the Fourier-transformed object function F{O} which is modulated with sin $\chi$ or cos $\chi$ 
\begin{equation}\label{eqn_in1}
F(I_{bright}) \propto F(O) sin \chi
\end{equation}
\begin{equation}\label{eqn_in2}
F(I_{pp}) \propto F(O) cos \chi
\end{equation}
The complex summation of these images $I_{complex} = I_{bright} + i \cdot  I_{pp}$ yields the expression 
\begin{equation}\label{eqn_in3}
F(I_{complex}) \propto F(O) sin \chi+ i \cdot F(O) cos \chi = F(O) exp(i \chi)
\end{equation}
The object information O can be retrieved dividing the Fourier-transform by the complex CTF and an inverse Fourier transformation.

\subsection{High-Resolution TEM with a Phase Plate}

\begin{figure}
\includegraphics[width=0.48\textwidth]{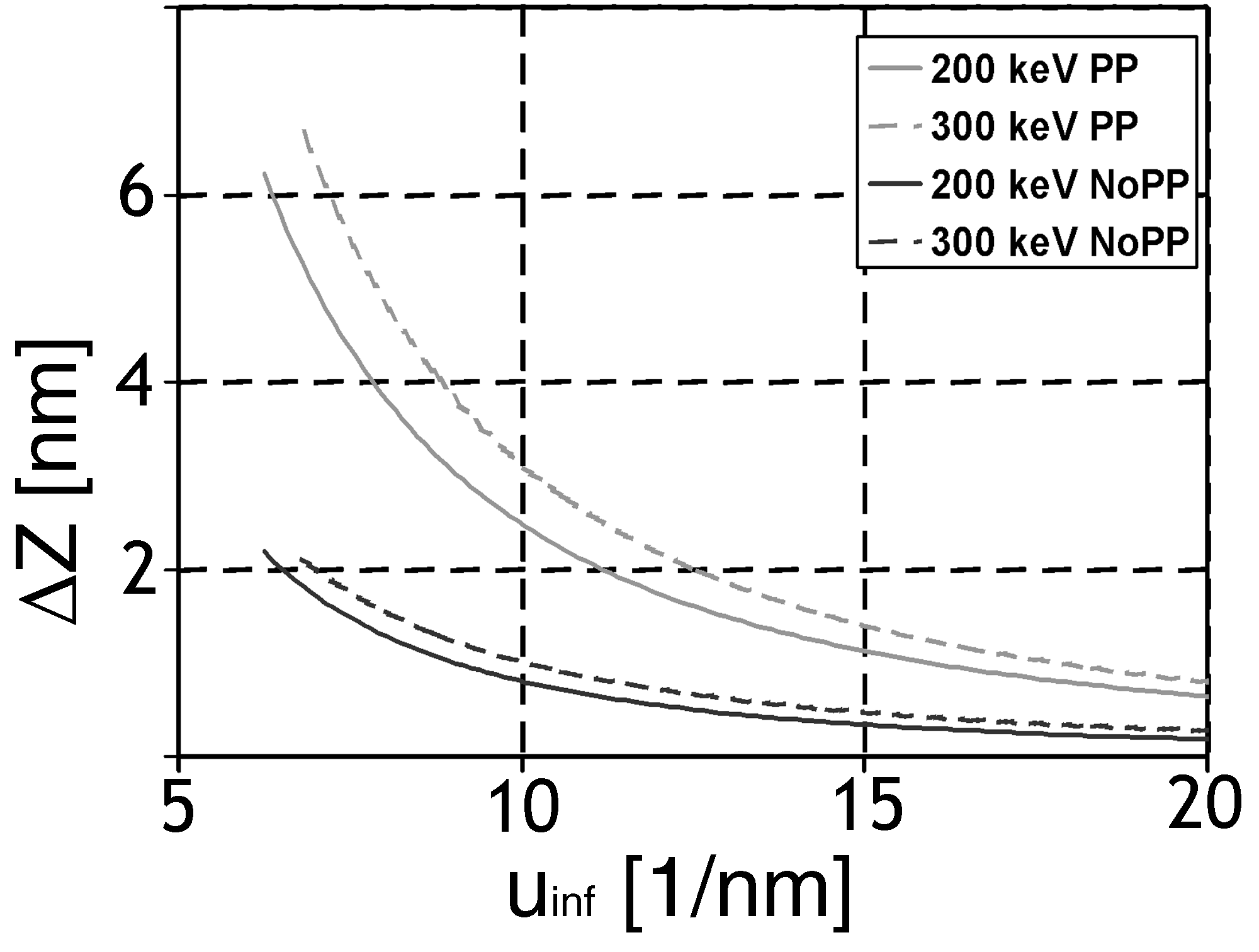}
\caption{\label{fig:fig4}Defocus interval with less than 5 \% reduction of the phase contrast integral PC (Eq. (\ref{eqn10})) with respect to the ideal case PC$_{id}$ (Eq. (\ref{eqn11})) for a PP microscope and a 5 \% reduction with respect to PC$_{len}$ for a non PP microscope plotted as a function of the information limit with phase plate (gray lines) and comparison to the case without phase plate (black lines).}
\end{figure}

While the required tolerances $\Delta C_{S}$ are easily met, the value of $\Delta Z$ may be limiting for practical work. Whether optimized conditions can be achieved, depends on the precision at which Z can be adjusted. This will not be a limitation for microscope 1 because the interval $\Delta Z  \approx 4~ nm$ for a 5 \% reduction of PC well exceeds the focusing accuracy. We can also assume that a future high-end microscope 2 will provide adequate focusing precision and stability. However, $\Delta Z$ also limits the field of view due to the surface topography of the specimen, which can be studied under optimized conditions. Fig. \ref{fig:fig4} shows the tolerances $\Delta Z$ as a function of $u_{inf}$ for microscope 1 and 2 with and without PP. $\Delta Z$ decreases strongly reaching values below 1 nm at $u_{inf} > 10~nm^{-1}$ for a non-PP microscope. This demands a high perfection of the sample-surface topography, which will strongly limit the field of view to be imaged under optimized conditions. The respective $u_{inf}$ for a PP microscope is well beyond $15~nm^{-1}$ permitting a larger field of view and rougher sample-surface topography at same resolution. In both cases microscope 2 shows a small advantage compared to microscope 1, i.e. the parameter intervals slightly benefit from higher acceleration voltages.

Experimental data from phase plate imaging show, that the weak-lens approximation holds for Boersch phase plates with electrode diameters around $1.5~\mu m$ \cite{Majorovits2007}. Size reduction of the electrode structure is desirable because the range of spatial frequencies, which are not transmitted, could be reduced. However, for a smaller electrode diameter of $1.1~\mu m$ the phase shift for different electrode voltages revealed an $u^{2}$-dependent distribution. Such a phase shift can be interpreted according to Eq. (\ref{eqn6}) as an additional defocus resulting from a finite focal length of the Boersch phase plate. While the effect may be small for uncorrected microscopes, it has to be considered for the case of $C_{S}$-correction, which is sensitive to very small defocus changes. Fig. \ref{fig:fig4} shows that the defocus windows $\Delta Z$ is reduced with increasing information limit and even a small defocus change by the phase plate cannot be neglected any longer. However, the observed electron deflection by the ring electrode does not impede the use of a Boersch phase plate for high-resolution TEM. The effect can be compensated when calculating the optimum defocus of the objective lens by considering the additional defocus.

Until now the influence of C$_{5}$ was neglected in our considerations. With current C$_{S}$-corrected microscopes C$_{5}$, if uncompensated, is below 10 mm \cite{Hosokawa2003} and does not affect phase contrast up to an information limit of 0.07 nm \cite{Chang2006}. For future correctors, which will allow higher resolution, the influence of C$_{5}$ can no longer be neglected. The aberration function is then given by Eq. (\ref{eqn13})
\begin{equation}\label{eqn13}
\chi = 2 \pi (\frac{1}{2} Z u^{2} + \frac{1}{4} C_{S} \lambda^3 u^4 + \frac{1}{6} C_{5} \lambda^5 u^6)
\end{equation}
To estimate the influence of C$_{5}$ in phase plate application, we use the phase contrast integral Eq.(\ref{eqn10}). We find that the influence is negligible up to an information limit of 0.083 nm (i.e. less then 1 \% change of the integral with C$_{5} = 10~mm$). This means that imaging with a phase plate is slightly more sensitive to C$_{5}$ than without a phase plate. For microscope 2 we assumed that C$_{5}$ is corrected to a value below 0.2 mm.

All preceding calculations implied $\phi_{pp} = + \pi/2$. Since the phase shift induced by the phase plate dominates the overall phase shift at small C$_{S}$- and Z-values according to Eq.(\ref{eqn6}), a positive PCTF is obtained in contrast to a negative PCTF for $C_{S} > 0$ and Z < 0 without phase plate (see Figs. \ref{fig:fig1}(a,b)). With respect to the work of Jia et al. \cite{Jia2004}, imaging with a PP microscope and $\phi_{pp} = \pi/2$ would correspond to negative C$_{S}$-imaging with bright atom contrast whereas $\phi_{pp} = - \pi/2$ would correspond to dark atom contrast. This situation can be realized with an electrostatic phase plate if the transmitted electrons are decelerated with respect to the scattered electrons. The choice between negative and positive phase shift is inherent to electrostatic phase plates while it is not for a Nagayama-type Zernike phase plate, which accelerates the electrons due to the positive mean inner potential of the amorphous carbon film. We therefore would expect bright atom contrast for a Zernicke phase plate where the scattered electrons are accelerated by the mean inner potential of the amorphous carbon film. 


\subsection{Obstruction of information due to phase plate structure}
The Boersch phase plate requires a ring-electrode structure fixed by supporting bars, which leads to obstruction of information. Information loss can be reduced, if the electrode is supported by bars, which are arranged in a three-fold symmetry. It was shown by Majorovits et al. \cite{Majorovits2007} that this particular geometry provides optimized single-sideband transfer for spatial frequencies otherwise obstructed by the supporting bars. However, the electrode structure cannot be removed. Even for the Zernike phase plate, which consists of an amorphous carbon film with a hole in its center, the electrons representing information at very low spatial frequencies - and which therefore pass the phase shifting film in its central hole - do not experience a phase shift. The range of these spatial frequencies is determined by the diameter of the transmitted beam, the wavelength and the focal length of the objective lens. We can relate the spatial frequency u in Fourier space with the radius r in the back focal plane in real space and the focal length f of the objective lens by Eq. (\ref{eqn14})
\begin{equation}\label{eqn14}
u = \frac{r}{\lambda f}
\end{equation}
Recent prototypes of such a Boersch phase plate fabricated by our group show an inner diameter of approximately $1~\mu m$ and outer diameter of the electrode ring of $2.7~ \mu m$. For a typical 200 keV microscope with a focal length of the objective lens f = 3 mm such a geometry will block information starting at the spatial frequency $u_{e}=0.18~nm^{-1}$  (i.e. object features larger in size than 5.56 nm). Electrons passing through the inner hole will not be enhanced in phase contrast and therefore will not transfer much information if the specimen is a weak-phase object. The amount of PC blocked by the electrode is 8\% for microscope 1, while the percentage decreases with increasing information limit. In the case of microscope 2 it is less then 3 \%. 
Although further reduction of the electrode structure is possible, there will be limits for the outer diameter of the electrode. Another route to minimize obstruction is the magnification of the back focal plane by a transfer doublet and the positioning of the phase plate in the magnified diffraction plane. This will extend effectively the focal length of the objective lens by a factor M = 5-10, and therefore the same electrode will only block information corresponding to object features larger than 28-56 nm.

\section{Conclusions}
A physical phase plate inserted into the back focal plane of a transmission electron microscope will allow superior phase contrast for weak-phase objects compared to conventional transmission electron microscopes. With suitable microscope parameters it is possible to obtain almost perfect coincidence of the phase contrast transfer function with the envelope function for partial temporal coherence even for future highly aberration-corrected microscopes. Ranges of optimal C$_{S}$- and Z-values are given for the best possible transfer, which depend on the information limit of the microscope. As a result of the small values of C$_{S}$ and the related small defocuses (in-focus phase contrast) a negligible contrast delocalization is achieved. 
In contrast to the conventional phase contrast imaging the in-focus phase contrast will provide strong and localized object contrast over the entire resolution range directly in one single image of the sample. This is most important for beam-sensitive samples, such as native biological samples under cryo-conditions, but will be of similar importance for materials science e.g. in the case of polymers. Other high-contrast / high-resolution imaging techniques, e.g. through focus series, need in general a larger electron dose and may therefore not be applicable. It should also be noted that the effective field of interpretable view increases for in-focus phase contrast imaging: Delocalization of object information in images of a through focus series cannot be compensated numerically at the border of the recorded images, i.e. a non-interpretable area remains at the edge of the image with a width, which is typically of the order of the contrast delocalization due to defocusing. Strong phase contrast at very low spatial frequencies and simultaneous high-resolution imaging over a large field of view will therefore better be obtained with phase plate mediated in-focus phase contrast.
While C$_{5}$ has limited effect on phase contrast for today’s microscopes, it will become important to correct its value, especially when using a physical phase plate in a microscope with an information limit below 0.083 nm. The imaging of samples with a physical phase plate will also allow high-resolution imaging for a wider spread of defocus, i.e. rougher sample-surface topography. To take full advantage of phase contrast imaging by means of a Boersch phase plate a further size reduction of the phase plate structure or a magnification of the back focal plane is required.
An interesting - and so far not conclusively studied - effect is the phase contrast of objects, which are strongly interacting with the electron beam and show large phase shifts of scattered electrons. Off-axis electron holography has been the state-of-the-art method to visualize such large phase shifting potentials, since it allows direct measurements of - in principle - arbitrarily large phase shifts. Physical phase plates in contrast rely in their basic mechanism on small relative phase shifts between scattered and unscattered electrons. Nevertheless, recently available in-focus phase contrast images of thick objects \cite{Nagayama2007,Barton2007} demonstrate that physical phase plates improve object contrast and signal localization for these samples as well. Further theoretical and experimental work will be necessary to describe these findings in detail.

\begin{acknowledgments}
The authors would like to thank B. Barton for stimulating discussions. The project is funded by the German Research Foundation (Deutsche Forschungsgemeinschaft) under Ge 841/16 and Sch 424/11.
\end{acknowledgments}

\bibliography{diss_main.bib}

\end{document}